\documentclass[aps,prb,floatfix,twocolumn,showpacs]{revtex4}
\usepackage{latexsym}
\usepackage{graphicx}
\newcommand{\figwidth}{3.275 in}
\usepackage{subfigure}
\usepackage{epsfig}
\begin{document}
\title{Effective Hamiltonian for FeAs based superconductors}
\author{Efstratios Manousakis$^{(1,2)}$, Jun Ren$^{(3)}$, Sheng Meng$^{(3)}$
and Efthimios Kaxiras$^{(3)}$}
\affiliation{
$^{(1)}$Department of Physics and MARTECH,
Florida State University,
Tallahassee, FL 32306-4350, USA\\
$^{(2)}$Department of  Physics, University of Athens,
Panepistimioupolis, Zografos, 157 84 Athens, Greece\\
$^{(3)}$Department of Physics and School of Engineering and
Applied Sciences, Harvard University,
Cambridge, MA 02138, USA
}
\date{\today}
\begin{abstract}
The recently discovered FeAs-based superconductors show intriguing
behavior and unusual dynamics of electrons and holes which
occupy the Fe $d$-orbitals and As $4s$ and $4p$ orbitals.
Starting from the atomic limit, we carry out a strong coupling expansion
to derive an effective hamiltonian that describes the electron
and hole behavior.
The hopping and the hybridization parameters between the
Fe $d$ and As $s$ and $p$-orbitals are obtained by fitting the
results of our density-functional-theory calculations to a tight-binding model
with nearest-neighbor interactions and a minimal orbital basis.
We find that the effective hamiltonian, in the strong on-site Coulomb
repulsion limit, operates on
three distinct sub-spaces coupled through
Hund's rule.  The three sub-spaces describe different components
(or subsystems): (a) one spanned by the $d_{x^2-y^2}$ Fe orbital; 
(b) one spanned by the degenerate atomic Fe orbitals $d_{xz}$ 
and $d_{yz}$;  and (c)
one spanned by the atomic Fe orbitals $d_{xy}$ and $d_{z^2}$.
 Each of these Hamiltonians is an extended $t-t^{\prime}-J-J^{\prime}$
model and is characterized by different
coupling constants and filling factors.  For the case of the undoped material
the second subspace alone prefers a ground state characterized by
a spin-density-wave order similar to that observed in recent experimental
studies, while the other two subspaces prefer
an antiferromagnetic order. We argue that the observed
spin-density-wave order minimizes the ground state energy of
the total hamiltonian.
\end{abstract}
\maketitle

\section{Introduction}

The recent observation of superconductivity in quaternary
oxypnictides\cite{JACSFeP,JACSFeAs,Ren,Chen1,Chen2,Wen},
which are materials based on FeAs (and FeP) ,
has rekindled intense activity\cite{Singh,Mazin,Hirschfeld,Kotliar,Boeri,Kotliar2,
Yu,Kuroki,Yildirim,AF,Yao,Qi,Scalapino,PALee,Tesanovic,Sachdev,Weng,Li}
to find a description of the strong electronic correlations present in these
materials and in the cuprates which could be responsible for such phenomena.

\begin{figure}[htp]
\vskip 0.4 in
\includegraphics[width=\figwidth]{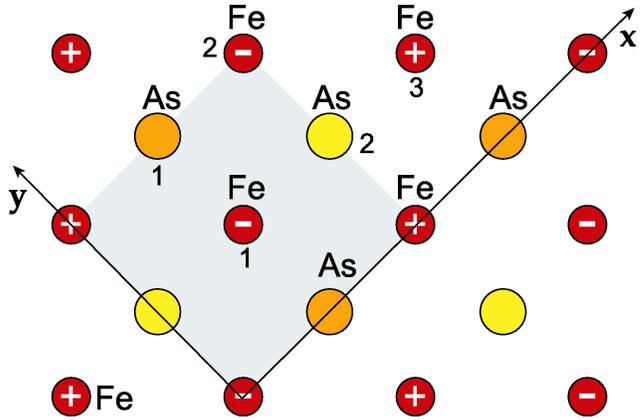}
\caption{The structure of the FeAs layer in the FeAs based superconductors.
The Fe atoms form a square lattice while the As atoms form two square 
sublattices
one just above (orange) and the other just below (yellow) the Fe plane. The
{\bf x} and {\bf y} axes used to characterize the orbitals are shown and the
unit cell is indicated by the shaded square.
The $+$ and $-$ signs
on the Fe atoms denote the spin orientation observed in the neutron
scattering experiment\cite{Cruz}.
The labels are used in the text to explain the
various types of antiferromagnetic exchange interactions.}
\label{fig1}
\vskip 0.2 in
\end{figure}

The structure of the new materials\cite{JACSFeAs} (Fig.~\ref{fig1}) consists of
FeAs layers sandwiched between LaO layers with rather weak
interlayer coupling.  The FeAs layers consist of a square lattice
formed by the Fe atoms, while the As atoms, which sit just above and
just below the plane (see Fig.~\ref{fig1}), form FeAs$_4$ octahedra squeezed
along the $c$ axis such that each Fe-As bond forms a $\pm 30^{\circ}$ angle
with the Fe plane.

As a function of temperature, the resistivity of the undoped parent
compound, which is not an insulator, shows a drop around 150 K\cite{JACSFeAs,Dong}
before turning back up below 50 K. In addition, the magnetic susceptibility
also shows an anomaly at 150 K and it was argued\cite{Dong} that
the parent material has a spin-density-wave (SDW) instability below
150 K. Recent neutron diffraction studies\cite{Cruz} demonstrate
that the parent compound at 150 K undergoes a structural distortion
from tetragonal at high temperature to monoclinic at low
temperature. Furthermore, these neutron diffraction studies show
that below $\sim 134$ K, while in the monoclinic phase, it develops
the SDW order shown in Fig.~\ref{fig1}. Subsequent M\"ossbauer
and muon spin rotation ($\mu SR$) studies\cite{Klauss} confirmed these findings:
the structural transition is found to be around 156 K and
the magnetic transition around 138 K. The structural
distortion which brings about the monoclinic structure at low
temperature is such that the rows of atoms which have their spins
antiferromagnetically aligned are closer than the rows of
atoms in the perpendicular direction.

The electronic structure of LaO$_{1-x}$F$_x$FeAs has been studied by
density functional theory\cite{Singh,Hirschfeld} and by
dynamical mean field theory\cite{Kotliar}.
There are arguments against phonon-mediated
superconductivity\cite{Kotliar,Boeri} in LaO$_{1-x}$F$_x$FeAs. In addition,
there are suggestions that a two-band
model\cite{Yu,Kuroki,Yao,Qi,Scalapino} may be the right effective hamiltonian
to use in order to describe the low energy physics of these materials.

Since the interlayer coupling is found to be
weak\cite{Hirschfeld}, in order to provide a simpler
basis to understand the electronic structure of these materials
we will focus on a single FeAs layer.
As further justification of this choice,  we have
calculated the band structure of LaOFeAs and that of a single FeAs
layer within density-functional theory and we find that
the important features of the bands near the Fermi level obtained
by the two calculations are essentially the same, including position relative
to the Fermi level and overall dispersion features.
Accordingly, our starting point is a hamiltonian which includes
the five Fe $d$-orbitals and four outer As orbitals ($4s$ and $4p$).
In our interacting electron model 
we include direct Fe-Fe hopping, hybridization
between the Fe $d$ and the As $4s$ and $4p$ orbitals, 
the local Coulomb repulsion energy for adding an electron on any
of the Fe $d$ states via Hubbard-type terms in the hamiltonian, and
finally, coupling through Hund's rule.

We determine the relative energy of the
atomic orbitals as well as the hopping and the hybridization matrix elements
as follows:  We carry out a set of first-principles calculations
of the electronic structure of a single FeAs layer.  
Using as basis the above mentioned Fe and As states,
we fit the results of the first-principles calculations using the tight-binding
approximation to determine the values of the hopping matrix elements and the
on-site energy levels. We find that the energy difference between the
atomic orbitals is less than 1 eV while the Coulomb repulsion to add
two electrons on the same Fe $d$ orbital is assumed to be
significantly larger\cite{Kotliar2,Hirschfeld,Abrahams,Shorikov,Kurmaev,Miyake}.
We perform a strong coupling expansion
in which the hopping and hybridization terms are used as perturbation
and the unperturbed parts are the terms which correspond to the
Coulomb repulsion energies
for a pair of electrons placed on the same or different orbitals.
Even in the case where these class of materials do not fulfil
the requirement for a strong coupling expansion, the qualitative
results or trends suggested by such a systematic analysis may still be useful.
We systematically derive a low energy effective hamiltonian to describe
this multi-band system.

Through this analysis, we show that the relevant low energy degrees of freedom
can be described by two 2-``flavor'' subsystems, one in which
the two flavors correspond to electrons in the $d_{xy}$ and $d_{z^2}$ states
of the Fe atoms and another
in which the two flavors correspond to $d_{xz}$ and $d_{yz}$ Fe orbitals,
and a third 1-flavor subsystem corresponding to the $d_{x^2-y^2}$ orbital.
Each of those subsystems is described by a
$t-t^{\prime}-J-J^{\prime}$ model where the spin degrees of freedom
of all three subsystems couple through Hund's rule.
Using this hamiltonian we are able to explain why the undoped material
orders in the SDW pattern shown in Fig.~\ref{fig1} as
reported by recent neutron scattering experiments\cite{Cruz}.

The paper is organized as follows: Section II gives a detailed discussion of
our density-functional-theory band structure calculations.  Section III presents
the strong-coupling limit analysis which leads to the effective spin-spin 
interaction hamiltonian derived in Section IV.
Section V gives a discussion of the physics of the effective hamiltonian.

\section{Electronic Structure Calculations}
\label{LDA}

\subsection{First-principles electronic structure}

\begin{figure*}[htp]
\includegraphics[width=5 in]{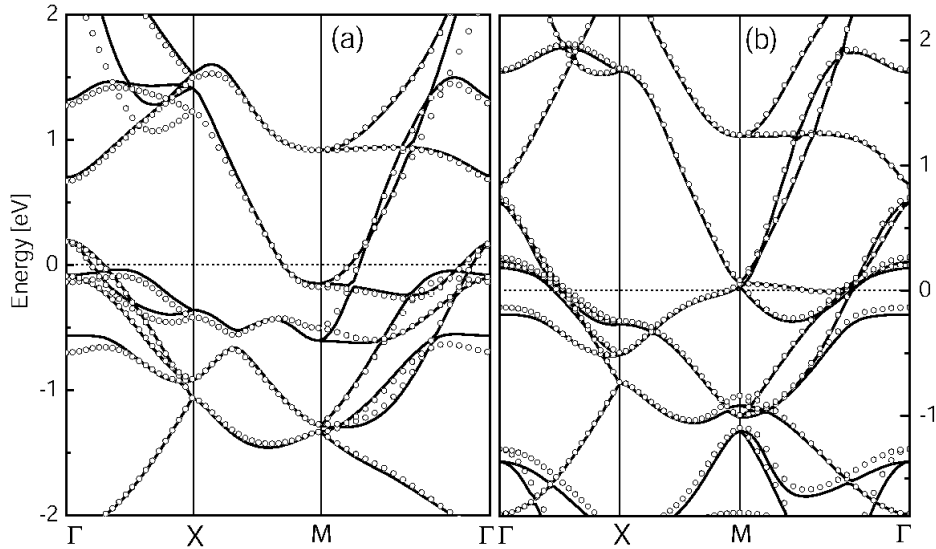}
\caption{The band structure obtained from the LDA calculations, 
for the paramagnetic phase of: (a) the LaOFeAs compound; (b) the FeAs layer.
In both cases the lines are from the SIESTA calculations and the points from 
the VASP calculation.  The Fermi level is set at zero in each case, for the 
neutral material.
}
\label{fig2}
\end{figure*}
Our first-principles calculations are performed within the
framework of density functional theory (DFT) and 
the local density approximation (LDA) 
for exchange-correlation effects.  We used the SIESTA
code~\cite{siesta} because 
it employs a localized basis of atomic-like orbitals for the expansion
of the wavefunctions which makes the interpretation of electronic
wavefunctions in the solid straightforward and transparent, without
the need for additional analysis such as projection to localized
Wannier-type orbitals. We use pseudopotentials of the
Troullier-Martins type~\cite{tro91} to represent the interaction
between valence electrons and ionic cores, and the Ceperley-Alder
form for the exchange-correlation functional~\cite{cep80}.
We have generated several different Fe pseudopotentials
in order to check for any dependence of the results on this ingredient 
of the calculations.  We find that the effect of the different 
pseudopotentials on the quantities reported 
in  the following, such as the electronic bands, is indiscernible.
In addition to the SIESTA calculations, we have used the VASP code
to ensure that there is reasonable agreement between the
results of two very different computational schemes.  The VASP
code uses a plane-wave basis instead of localized orbitals \cite{VASP} and
employs pseudopotentials of a different type \cite{VASP-PSP} than those
in the SIESTA calculation.

LaOFeAs, belonging to the tetragonal P4/nmm space group, has a
layered structure~\cite{JACSFeAs}. The FeAs layer serves as the
carrier conduction channel and it has strong electronic
couplings within the layer.  The unit cell for
the simplified model system, a single FeAs layer, contains two Fe
and two As atoms with a vacuum layer with thickness $\sim 19$~\AA.
The full system, bulk LaOFeAs has two atoms of each type (Fe, As, La and O)
in the unit cell. 

In the SIESTA calculations, we choose an auxiliary real space
grid equivalent to a plane-wave cutoff of 100 Ry, and use
$8\times8\times1$ Monkhorst-Pack k-point grid for the FeAs layer,
$4\times4\times1$ for the (2$\times$2) FeAs supercell, and a
$8\times8\times4$ grid for the LaOFeAs structure. For
geometry optimization, a structure is considered fully relaxed
when the magnitude of forces on all atoms is smaller than 0.04
eV/\AA.  In the VASP calculation we use the same k-point grids and 
the default plane-wave cutoffs.

\begin{figure}[htp]
\includegraphics[width=3.5 in]{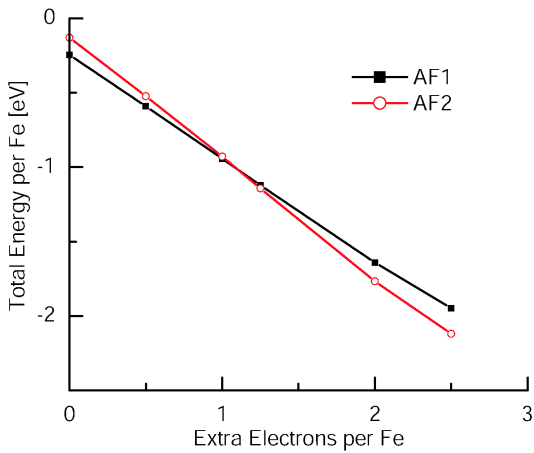}
\caption{
The total energy of the FeAs layer as a function of the added charge
obtained with the VASP code.  The reference energy is the energy of 
the neutral layer in the paramagnetic phase.  
}\label{fig3}
\end{figure}

The results obtained for bulk LaOFeAs and for the FeAs layer, 
using the SIESTA and VASP codes,
for the paramagnetic phase are shown in Fig. \ref{fig2}.
In both cases, the agreement between the two different computational 
schemes is remarkable.  Moreover, there a very close similarity between the
bands of the bulk LaOFeAs compound and the FeAs layer, 
especially in what concerns the features near the Fermi level.
Notice that the neutral FeAs layer contains a different number of 
electrons per Fe atom than the bulk LaOFeAs compound.  
Accordingly, we show the Fermi levels for the bulk and the layer
slightly offset, to emphasize the similarity of the band-structure features.
Moreover, by examining carefully the wavefunction character 
of the bands near $\Gamma$ and M, the points in the Brillouin Zone
where electron and hole pockets appear in the bulk LaOFeAs 
compound, we establish that these states arise from orbitals
associated with the Fe and As atoms.  For these reasons, 
it is reasonable to concentrate on the band-structure of the 
FeAs layer alone, in order to build a comprehensive picture 
of the interacting electron system, presented in the following sections.

We address next the issue of the spin configuration.
We considered different spin configurations in the anti-ferromagnetic 
(AFM) phase. For these calculations we use a (2$\times$2) FeAs supercell.
The two spin configurations are: (i) AFM1, which is simply repetition
of the spin configuration in the (1$\times$1) unit cell; (ii)
AFM2, which has the same spin alignment in one of the two diagonal
directions of the Fe lattice and alternating spins along the other
direction, as shown in Fig.~\ref{fig1}.
The total energy for the former spin configuration 
is lower by 0.95 eV per $(2\times 2)$ cell,
suggesting the AFM1 configuration is more stable than the AFM2 one
for the neutral FeAs layer.
We believe that this result is due to the fact that our calculation is for the
charge neutral FeAs layer, where the Fermi level is lower than that
of the LaOFeAs compound;  namely, the FeAs layer as part of the
 LaOFeAs structure is negatively charged by an extra electron per 
Fe atom because the LaO layer is positively charged since
the preferred oxidation state of La is La$^{3+}$ and that 
of oxygen is O$^{2-}$.
To show that this is the case, we carried out a calculation
using the VASP code for a charged FeAs layer. In Fig.~\ref{fig3}
the total energies of the AFM1 and AFM2 phases are compared
as a function of the added charge $\sigma$ per Fe atom.
Notice that the AFM2 phase
becomes energetically favorable for $\sigma \sim 1$.  
The exact values of $\sigma$ can not be
accurately determined in the context of these 
calculations because the energy differences are within the range 
of accuracy of DFT-LDA.
The fact that the AFM2 phase becomes the ground state for
$\sigma =1 $ will be established using the effective
Hamiltonian derived in the next section, which captures the interacting-electron
nature of the system in a more realistic manner, and as such, gives
more reliable results for the magnetic phases.
\bigskip

\subsection{Tight-binding approximation model}
\begin{figure}[htp]
\includegraphics[width=3.275 in]{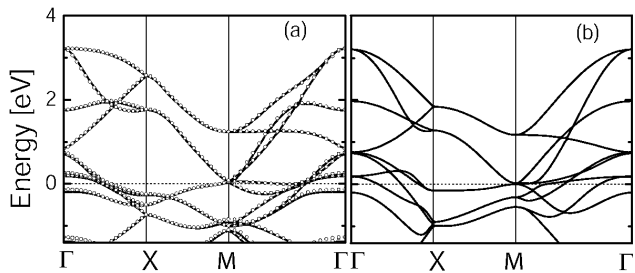}
\caption{The band structure of FeAs layer in the paramagnetic
phase  obtained from: (a) the LDA calculations with the two
different approaches (SIESTA - lines, and VASP - points);
(b) the tight-binding approximation with nearest neighbor interactions
and minimal orbital basis.
}\label{fig4}
\end{figure}

Our goal next is to calculate a tight-binding hamiltonian which approximately
gives the same band structure as that of the first-principles results for 
the FeAs layer, using nearest-neighbor interactions only
(hopping matrix elements) and a minimal orbital basis, 
consisting of the $4s$ and $4p$ As orbitals and the $3d$ Fe orbitals.  
We expect that the hopping matrix elements, needed for the
hamiltonian upon which the strong coupling expansion will be based,
are not significantly affected by the value of the
filling factor (the position of the Fermi level). 
For the reasons discussed above, namely that a realistic picture 
of spin configurations can only arise from the interacting-electron treatment
based on the effective hamiltonian, we will focus on 
reproducing with the tight-binding approximation 
the band structure of the paramagnetic phase, as 
obtained from the DFT-LDA calculations. 

Since introduction of electron doping
is necessary in order to produce superconductivity in LaOFeAs-based
materials, we focus in reproducing as accurately as possible the 
features near and above the Fermi level.
Note that hole-doping induced
superconductivity in these materials has been also reported recently\cite{Wen}.
We also use information from the first-principles electronic wavefunctions
to determine what is the optimal fit of the tight-binding approximation 
to the DFT-LDA results. As mentioned above, all the bands
in the neighborhood of the Fermi level are associated with the 
As $4s$ and $4p$ and the Fe $3d$ orbitals, and these features are 
well reproduced by the tight-binding approximation results.
The best fit we could achieve is shown in Fig.~\ref{fig4}, and compared 
to the first-principles results for an extended region near the Fermi level.
The on-site and hopping matrix elements that produce this
fit are presented in the next section, where these values are employed 
to construct the effective hamiltonian. 

\section{Strong Coupling Limit}
\label{strong}

We consider the hamiltonian describing a single FeAs layer of
Fe and As (or, in a more general formulation, P) atoms:
\begin{eqnarray}
\hat H = \hat H_a + \hat T + \hat U,
\end{eqnarray}
with the three terms defined by:
\begin{widetext}
\begin{eqnarray}
\hat H_a &=& \sum_{i,\nu,\sigma} \epsilon_d(\nu) d^{(\nu)\dagger}_{i\sigma}
d^{(\nu)}_{i\sigma}  + \sum_{l,\alpha,\sigma} \epsilon_{sp}(\alpha)
sp^{(\alpha)\dagger}_{l\sigma} sp^{(\alpha)}_{l\sigma}, \\
\hat T &=& - \sum_{<ij>,\sigma,\nu\nu'} (t_{\nu\nu'}
d^{(\nu')\dagger}_{j\sigma}d^{(\nu)}_{i\sigma} + h.c)
- \sum_{i,\sigma,\nu,\alpha}
\sum_{ l(i)} (V^{\nu\alpha}_{il} d^{(\nu)\dagger}_{i\sigma}
sp^{(\alpha)}_{l\sigma}  + h.c),   \\
\hat U &=& \sum_{\nu}U^{(d)}_{\nu} \sum_i
n^{\nu}_d(i\uparrow)n^{\nu}_d(i\downarrow)
 + \sum_{\nu,\nu'\ne\nu}U^{(d)}_{\nu\nu'} \sum_{i\sigma\sigma'}
n^{\nu}_d(i\sigma)n^{\nu'}_d(i\sigma') \nonumber \\
&+& \sum_{\alpha}U^{(sp)}_{\alpha} \sum_i
n^{\alpha}_{sp}(i\uparrow)n^{\alpha}_{sp}(i\downarrow)
+ \sum_{\alpha,\alpha'\ne\alpha}U^{(sp)}_{\alpha\alpha'} \sum_{i \sigma\sigma'}
n^{\alpha}_{sp}(i\sigma) n^{\alpha'}_{sp}(i\sigma')
- \sum_{\nu,\nu'}J^H_{\nu,\nu'} \sum_i {\vec S}^{\nu}_d(i) \cdot
{\vec S}^{\nu'}_d(i).
\end{eqnarray}
\end{widetext}

We discuss the non-interacting parts, $\hat{H}_a$ and $\hat{T}$ of
the hamiltonian first.
The operator $d^{(\nu)\dagger}_{i\sigma}$ creates an electron
of spin $\sigma$ on the $\nu^{th}$ Fe $d$-orbital
($\nu=1,2,3,4,5$ is the index that corresponds to the five $d$ Fe
orbitals, $d_{x^2-y^2}$, $d_{xz}$, $d_{yz}$, $d_{xy}$,  and $d_{z^2}$
respectively)
at the site $i$ which has an energy $\epsilon_d(\nu)$.
The operator $sp^{(\alpha)\dagger}_{l\sigma}$ creates
an electron of spin $\sigma$ on the $\alpha^{th}$
As which is one of {\it three}  As orbitals.
These As orbitals are formed as follows: first, because of the tetragonal
symmetry, the two $4p_x$ and $4p_y$ 
orbitals remain unhybridized, while the $4s$ and $4p_z$ As states form
two linear combinations 
$|sp^{\pm}_z\rangle = a |s\rangle \pm b |p_z\rangle$.
The LDA calculation shows that the As $sp^-_z$ state together
with the Fe $4s$  form a bonding and an anti-bonding  state, with
the bonding state approximately $ 10$ eV below the Fermi level
and the anti-bonding state approximately $ 6$ eV above the
Fermi level. Therefore, these two states are not included
in the tight-binding fit and the three As states included are the
 $4p_x$ and $4p_y$ orbitals and the $sp^+_z$ hybrid.
Thus, $\alpha=1,2,3$ corresponds to 
the cases of $4p_x$,$4p_y$ and $sp^+_z$ respectively
at the $l^{th}$ As site, with site energy
 $\epsilon_{sp}(\alpha)$. 
 
 $V$ is the hybridization term
between the Fe $3d$ orbitals and the As orbitals.
The sum over $l(i)$ means that it is over
all four As sites $l$ around the $i^{th}$ Fe site. The hybridization
matrix element $V^{\nu\alpha}_{il}$ is proportional to the wave function
overlap of the $\nu^{th}$ Fe $d$-orbital and the As $\alpha^{th}$
$sp$-orbital. Some of these matrix elements are zero due to symmetry
arguments and the most significant ones are of the order of, or less
than, 1 eV, as obtained through our tight binding fit of the
LDA results.

\begin{table}
\caption{The on-site energies in eV for the Fe $3d$ orbitals
as determined by approximating the results of the first-principles
band-structure calculation, using the
tight-binding approximation discussed in Sec.~\ref{LDA}.  We also include the
hopping matrix elements $t_{\nu\nu}$ between two nearest
neighbor Fe $d$-orbitals of the same type.
The notation is explained in Sec.~\ref{strong}.}
\label{table1}
\begin{tabular}{|p{60pt}|p{30pt}| p{30pt}|p{30pt}|p{30pt}|p{30pt}|}\hline
$\nu$ & 1 & 2 & 3 & 4 & 5 \\ \hline
Fe $3d$-orbital& $d_{x^2-y^2}$ & $d_{xz}$ & $d_{yz}$ &$d_{xy}$ & $d_{z^2}$ \\ \hline
$\epsilon_d(\nu)$  & $-4.6 $  & $-4.5 $ & $-4.5 $
&$ -4.5$  & $-4.2 $ \\ \hline
$t_{\nu\nu}$  & 0.22  & 0.5  & 0.5  & 0.43  & 0.22  \\ \hline
\end{tabular}
\end{table}

\begin{table}
\caption{Same as in Table \ref{table1}, for the As $4s4p$-orbitals.
}
\label{table2}
\begin{tabular}{|p{60pt}|p{30pt}| p{30pt}|p{30pt}|}\hline
$\alpha$ & 1 & 2 & 3  \\ \hline 
As sp-orbital & $p_x$ & $p_y$ & $sp^+_z$  \\ 
\hline $\epsilon_{sp}(\alpha)$  & $-4.7 $ & $-4.7 $ & $-5.2$ 
\\ \hline
\end{tabular}
\end{table}

\begin{table}
\caption{Same as in Table \ref{table1}, for 
the hybridization matrix elements between Fe $3d$- and
As $4s4p$-orbitals.
The atom labels are those shown in Fig.~\ref{fig1}.
For the case of Fe atom labeled
2 in Fig.~\ref{fig1} the matrix elements are obtained from the
same table by interchanging the labels 1 and 2 of the
As orbitals and reversing the sign. }
\label{table3}
\begin{tabular}{|p{25pt}|p{40pt}| p{40pt}|p{40pt}|p{40pt}|p{40pt}|}\hline
$V^{\nu\alpha}$& $1(d_{x^2-y^2})$  & $2(d_{xz})$ & $3(d_{yz})$ & $
4(d_{xy})$ & $5(d_{z^2})$\\ \hline 
$p_x(1) $   & 0    & 0.1  & 0        & $-0.2$    & 0 \\\hline 
$p_x(2) $   & 0.4    &$-1.45$ & 0     & 0  & 0.25 \\ \hline
$p_y(1)$    & $-0.4$    & 0    & 1.45  &   0 & 0.25\\\hline 
$p_y(2)$    &0  & 0    & $-0.1$  & $-0.2$    & 0 \\ \hline 
$sp^+_z(1)$ & $-0.5$    & 0    & 0.7   & 0 & 0.9  \\\hline
$sp^+_z(2)$ & $-0.5$   & 0.7  & 0     & 0 & $-0.9$  \\
\hline
\end{tabular}
\end{table}

In Tables~\ref{table1},\ref{table2} and in Table~\ref{table3} we
give the non-zero matrix elements obtained by fitting the LDA
results to the tight binding model (as explained in
Sec.~\ref{LDA}) which includes the five Fe $d$-states and the three
As $4s-4p$ states for each of the two Fe and the two As atoms in the
Fe$_2$As$_2$ unit cell, as well as the matrix elements $t_{\nu\nu'}$ and
$V^{\nu\alpha}$ between these states.  
In addition to the above terms, the tight-binding
approximation to the LDA results gives two hopping matrix elements
$t_{\nu\nu'}$: the first for $\nu \to d_{xz}$ and $\nu' \to d_{yz}$, which is
$t_{xz,yz}=0.54$ eV, and the second for $\nu \to d_{xy}$ and $\nu' \to d_{z^2}$,
which is $t_{xy,z^2}= 0.20 $ eV. All other hopping matrix elements
are either identically equal to zero due to symmetry or negligibly small.

We turn next to the interaction part, $\hat{U}$, of the hamiltonian.
$n^{\nu}_d(i\sigma)=
d^{(\nu)\dagger}_{i\sigma}d^{(\nu)}_{i\sigma}$ is the number operator
and $U^{(d)}_{\nu}$ or $U^{(sp)}_{\alpha}$ give the Coulomb repulsion 
for a pair of electrons
placed on the same $d$-orbital or the same $s$ or $p$ As-orbital.
The $U^{(d)}_{\nu\nu'}$ (or $U^{(sp)}_{\alpha\alpha'}$ )is responsible for
the Coulomb repulsion between
different Fe-$d$ (or As $sp$) orbitals within the same atom. We will assume
that the  Coulomb-repulsion terms between the same or different Fe-$d$
orbitals are significantly greater than their
counter-parts for the As  $sp$ states, consistent with the general expectations
for these values in the 
literature: $U^{(d)}$ is believed to be large of the order
of 4-5 eV\cite{Kotliar2,Hirschfeld,Abrahams,Shorikov,Kurmaev,Miyake}
 while the parameter $U^{(sp)}$ is expected to be much smaller than that.
In addition, we will assume that the same site Fe-$d$ Coulomb repulsion
is larger than the inter-orbital Coulomb repulsion
$U^{(d)}_{\nu\nu'}$. The term proportional
to $J^H$ represents Hund's rule for the Fe d orbitals, 
with $J^H>0$, of order of less than 1 eV. We have neglected the
Hund's rule coupling for  As orbitals.

\begin{figure}[htp]
\vskip 0.4 in
\includegraphics[width=\figwidth]{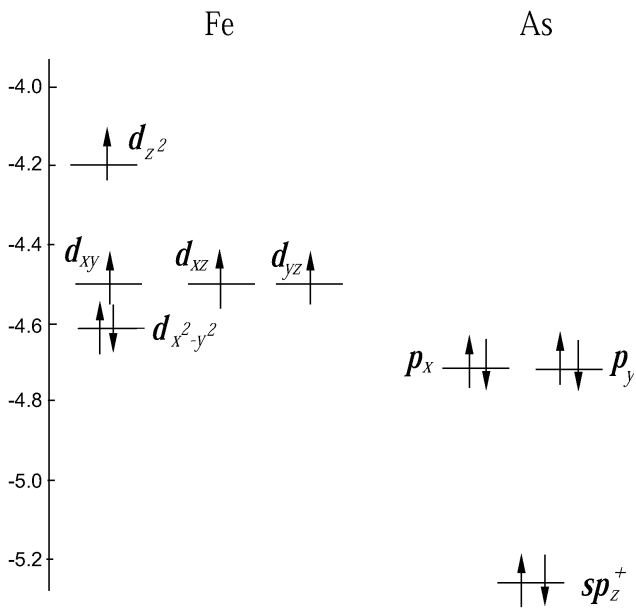}
\caption{The occupation of the Fe and As atomic levels
in the FeAs plane of the undoped parent compound.  The ordering of the
levels is shown schematically, as obtained from the tight-binding
approximation parameters.}
\label{fig5}
\vskip 0.2 in
\end{figure}

Notice that the energy levels $\epsilon_d(\nu)$ and $\epsilon_{sp}(\nu)$ lie
in the region $-4.7 \pm 0.5$ eV, namely the energy difference between
any pair of such states is less than  1 eV which is
believed to be smaller than the characteristic Coulomb repulsion 
energy $U^{(d)}$. In this paper we begin our analysis from the atomic
or strong coupling limit, which implies that we have assumed that
the energy scale $U^*$ defined below in Eq.~\ref{ustar} is significantly 
larger than the hopping and hybridization matrix elements. If 
this condition is not fulfilled for this class of materials, it may still
be instructive to discuss the qualitative features which a strong
coupling expansion yields.
In this limit the unperturbed part of the
hamiltonian $\hat H_0$ is
\begin{eqnarray}
\hat H_0 = \hat H_a + \hat U,
\end{eqnarray}
and the hopping part $\hat T$, which  includes the hybridization,
plays the role of perturbation.
As discussed previously, the FeAs layer in the undoped
LaOFeAs parent compound has an additional electron relative to
the neutral FeAs layer. This is due to the fact that LaO layer is expected
to be in combined $1+$ oxidation state. 
Therefore, the eight states considered above (five Fe $d$ states plus
three As $sp$ states) are occupied by 12 electrons.
The atomic configuration is shown in Fig.~\ref{fig5}. Since we have
five more electrons than levels, five energy levels must be
doubly occupied. Double occupancy of the Fe $d$ orbitals is 
much more costly 
compared to the As $sp$ orbitals; 
therefore, all three of the As orbitals must be doubly occupied
and the only doubly occupied Fe $d$ orbital is $d_{x^2-y^2}$, which has
the lowest energy. The other four Fe $d$ orbitals 
are singly occupied and the spin of these electrons
are parallel because of Hund's rule, represented by the coupling $J_H$.

Next we consider the effective hamiltonian which,
in strong coupling perturbation theory,  is given by
\begin{eqnarray}
\hat H_{eff} &=& E_0 \hat P_0
 + \hat P_0 \hat T \hat \Omega, \\
\hat \Omega &=& \hat P_0 + \hat R (\hat T \hat \Omega - \hat \Omega \hat T
\hat \Omega), \\
\hat R &=& {{\hat Q} \over {\hat H_0 - E_0}},
\end{eqnarray}
where $E_0$ is the ground state energy given by the energy of the
state depicted in Fig.~\ref{fig5}. The $\hat H_{eff}$
operates in the subspace $S_0$ formed
by the degenerate ground states of $\hat H_0$, that is, the
subspace of states produced by the direct product of atomic states like those in
Fig.~\ref{fig5} in which the spins of the four electrons occupying the four
$d$-orbitals, from one Fe atom to the next, point either all up or all down.
The operator $\hat P_0$ is a projection operator which projects into
the subspace $S_0$ and $\hat Q=\hat 1-\hat P_0$, that is, the
operator which projects outside the subspace $S_0$. The above equation
can be formally solved iteratively to yield the Rayleigh-Schr\"odinger
expansion as a power series in $\hat T$.
The leading term is $\hat T$ which, when restricted in this subspace $S_0$,
becomes just the direct Fe-Fe hopping terms
\begin{eqnarray}
\hat P_0 \hat T \hat P_0 = - \sum_{<ij>,\sigma,\nu\nu'} (t_{\nu\nu'}
d^{(\nu')\dagger}_{j\sigma} d^{(\nu)}_{i\sigma} + h.c),
\end{eqnarray}
In the right-hand-side of the above equation we have omitted the
projection operators by assuming that we will restrict ourselves to
the subspace $S_0$.

In the following,
in order to simplify the calculation, we will
take $U^{(d)}_{\nu}=U$ to be independent of $\nu$ and we will assume that
$U$ is much larger than the atomic energy level difference (which was
found to be less than 1 eV  within our TB approximation) and 
significantly larger than
the hopping and hybridization parameters.
In addition,  we will take $U^{(d)}_{\nu \ne \nu'}=\bar U$,
 $U^{(sp)}_{\nu}=U_{sp}$ and $U^{(sp)}_{\alpha \ne \alpha'}={\bar U}_{sp}$, i.e.,
to be independent of  $\nu, \nu'$ (or $\alpha, \alpha'$).

\section{Effective spin-spin interaction hamiltonian}

\subsection{Interaction between same-type Fe orbitals}

First, there are the familiar second order processes arising
from the direct Fe-Fe hopping through the matrix elements $t_{\nu\nu}$
which give rise to an antiferromagnetic exchange interaction of the form
\begin{eqnarray}
{\cal H}_{\nu,\nu} &=& J^{(2)}_{\nu,\nu} \sum_{<ij>}
\vec S^{\nu}_i \cdot \vec S^{\nu}_j,\label{AFsecond}\\
J^{(2)}_{\nu,\nu} &=& {{4 t_{\nu,\nu}^2} \over {{U}}},\label{AFsecondJ}.
\end{eqnarray}
In addition, we have antiferromagnetic coupling of the spins of two
different type Fe $d$-orbitals due to the hopping terms $t_{xz,yz}$ and
$t_{xy,z^2}$ which give rise to
\begin{eqnarray}
{\cal H}_{\nu,\nu'} &=& J^{(2)}_{\nu,\nu'} \sum_{<ij>}
\vec S^{\nu}_i \cdot \vec S^{\nu'}_j,\\
J^{(2)}_{\nu,\nu'} &=& {{4 t^2_{\nu,\nu'}} \over {U}},
\end{eqnarray}
where $\nu,\nu'$ can be either $xz,yz$ or $xy,z^2$.
These processes take place only between nearest neighbors such as the
Fe atoms 1 and 2 in Fig.~\ref{fig1}. These second order
contributions are obtained from the square of the matrix
elements listed in Table~\ref{table4} by multiplying them
with $4/U$. There are no second-order nnn contributions to the
spin-spin interaction.

\begin{table}
\caption{The second-order nn contribution to the 
spin-spin couplings is obtained from the matrix elements listed below
(in units of eV$^2$)
by multiplying them with $4/U$. 
}
\label{table4}
\begin{tabular}{|p{30pt}|p{30pt}| p{30pt}|p{30pt}|p{30pt}|p{30pt}|}\hline
$t^2_{\nu\nu'}$
& $d_{x^2-y^2}$ & $d_{xz}$ & $d_{yz}$ & $d_{xy}$  & $d_{z^2}$\\ 
\hline $d_{x^2-y^2}$ &  0.048  & 0  & 0  & 0 & 0 \\ 
\hline $d_{xz} $& 0  & 0.25 & 0.29 & 0 & 0  \\ 
\hline $d_{yz}$ & 0 & 0.29  & 0.25  & 0 & 0 \\
\hline $d_{xy} $& 0 & 0  & 0 & 0.185 & 0.04 \\
\hline $d_{z^2}$ & 0  & 0  & 0   & 0.04 & 0.048  \\\hline
\end{tabular}
\end{table}

The next terms to leading order, beyond the first and second order terms
discussed above, are fourth order processes involving an Fe $d$-orbital
and the As $sp$-orbitals with which it hybridizes.
The contribution of all $sp$ orbitals of the two possible intervening As
atoms to the exchange interaction between the same $d$ orbital
of two nn Fe atoms is given by\cite{Jefferson}
\begin{eqnarray}
{\cal H}_{\nu,\nu} &=& J^{(4)}_{\nu,\nu} \sum_{<ij>} 
\vec S^{(\nu)}_i \cdot \vec S^{(\nu)}_j,\\
J^{(4)}_{\nu,\nu} &=& \sum_{\alpha=1}^6 J^{\alpha}_{\nu\nu}, \\
J^{\alpha}_{\nu\nu} &=& 2 b^2 \Bigl [ { 1 \over {U^*}} +
{1 \over {U^*    + \epsilon}} \Bigr ], \hskip 0.2 in
 b = {{V^{\alpha\nu}_1 V^{\alpha\nu}_2} \over {U^* + \epsilon}},
\label{NNJ}
\end{eqnarray}
where
\begin{eqnarray}
U^* = U + 5 \bar U -(4 {\bar U}_{sp} + U_{sp}),
\label{ustar}
\end{eqnarray}
 and $U^*$ is assumed significantly larger that all other energy scales in the
problem. Here 
\begin{eqnarray}
\epsilon =\epsilon_{sp}({\alpha})-\epsilon_d(\nu),
\label{epsilon} 
\end{eqnarray}
is the energy difference between the
As $sp$ state and that of the Fe $d$ orbital. The subscripts 
1 and 2 in the matrix elements refer to the fact that
the two nn atoms 1 and 2 are at $90^{\circ}$ angle relative to
the position of the intervening As atom and, therefore,
$V_1$ is from Table~\ref{table3}, while $V_2$ from 
Table~\ref{table3} by interchanging the index 1 and 2 and the sign of
the matrix elements.

Since $\epsilon < 1$ eV for any combination of Fe-$d$ and the three As $sp$
orbitals and $U^* \sim 5$ eV, if $\epsilon$ is neglected 
in the above expression the error in our estimate for the
exchange couplings $J$ will be rather small. We are going to
use the full expression given by Eq.~\ref{NNJ} when we
compute the coupling constants $J$ between the same-type Fe orbitals. 
We will also use this approximation of neglecting $\epsilon$ because 
it simplifies the results and this gives additional insight. 
With this approximation, we find that
\begin{eqnarray}
J^{(4)}_{\nu,\nu} &\simeq& {{4 A_{\nu\nu}} \over {{U^*}^3}},\\
A_{\nu\nu} &=& \sum_{\alpha=1}^6  (V^{\alpha\nu}_1 V^{\alpha\nu}_2)^2.
\label{diagonalJ}
\end{eqnarray}
Since the value of $U^*$ is not known, we can use the expression
given by Eq.~\ref{diagonalJ} as a measure of the relative
spin-exchange interaction coupling. Namely, to
obtain the actual values of $J^{(4)}_{\nu\nu}$, and the 
$J^{(4)}_{\nu\nu'}$ with $\nu \ne \nu'$ to be discussed next, we just need to
multiply the values given in Table~\ref{table5} by $4/(U^*)^3$.
Therefore,  for two
nn Fe atoms, such as Fe atoms 1 and 2 in Fig.~\ref{fig1},
the spin-spin interaction coupling constant $J_{\nu\nu}$ is obtained
by adding the second and fourth order contributions
$J^{(2)}_{\nu\nu}$ and $J^{(4)}_{\nu\nu}$ respectively.

On the other hand, for two next nn Fe atoms,
 such as 1 and 3 in Fig.~\ref{fig1}, there are no second order
processes since there are no direct hopping between such atoms.
The fourth order superexchange contributions is obtained as follows:
\begin{eqnarray}
{\cal H}^{\prime}_{\nu,\nu}
&=& J^{\prime}_{\nu,\nu} \sum_{<<ij>>} \vec S^{(\nu)}_i \cdot \vec S^{(\nu)}_j,\\
J^{\prime}_{\nu,\nu} &=& {1 \over 2}\sum_{\alpha=1}^6 J^{\prime\alpha}_{\nu\nu}, \\
J^{\prime \alpha}_{\nu\nu} &=& b'^2 \Bigl [ { 1 \over {U^*}} +
{1 \over {U^*    + \epsilon}} \Bigr ], \hskip 0.2 in
 b^{\prime} = 2 {{(V^{\alpha\nu}_1)^2} \over {U^* + \epsilon}}.
\end{eqnarray}
Notice, that in this case the same matrix elements $V_1$ are
involved because both atoms participating in the superexchange form 
the same angle with the intervening As atoms.
In addition, there is a factor of 2 difference between the above
expression and Eq.~\ref{NNJ} because there is
only one possible intervening As atom for two fixed Fe atoms.
The simplified expression, when $\epsilon$ is neglected is the following:
\begin{eqnarray}
J'^{(4)}_{\nu,\nu} &\simeq& {{4 A'_{\nu\nu}} \over {{U^*}^3}},\\
A'_{\nu\nu} &=& {1 \over 2} \sum_{\alpha=1}^6  (V^{\alpha\nu}_1)^4.
\label{diagonalJp}
\end{eqnarray}
The values of the constant $A'_{\nu\nu}$ are given 
as the diagonal matrix elements of Table~\ref{table6}.

\subsection{Interaction between different-type Fe orbitals}

There is an effective spin-spin interaction between
certain Fe orbitals of different type, $J_{\nu \nu'}$ 
with $\nu\ne \nu'$.  This type of nn and nnn spin-spin interaction 
for most of the orbitals is
significantly smaller than the $J_{\nu\nu}$ coupling constants
between the same orbitals.
To obtain an estimate of these we use the simplified expressions
where $\epsilon$ is neglected. Namely,
\begin{eqnarray}
J^{(4)}_{\nu,\nu'} &\simeq& {{4 A_{\nu\nu'}} \over {{U^*}^3}},\\
A_{\nu\nu'} &=& \sum_{\alpha=1}^6  (V^{\alpha\nu}_1 V^{\alpha\nu'}_2)^2,\\
J'^{(4)}_{\nu,\nu'} &\simeq& {{4 A'_{\nu\nu'}} \over {{U^*}^3}},\\
A'_{\nu\nu'} &=& {1 \over 2} \sum_{\alpha=1}^6  (V^{\alpha\nu}_1
V^{\alpha\nu'}_1)^2.
\label{off-diagonal}
\end{eqnarray}
The values of the constants $A_{\nu\nu'}$ and $A'_{\nu\nu'}$ are given 
as the off-diagonal matrix elements of Tables~\ref{table5} and \ref{table6}.

\begin{table}
\caption{Fourth-order sum of matrix elements 
contributing to the nn spin-spin couplings.}
\label{table5}
\begin{tabular}{|p{30pt}|p{30pt}| p{30pt}|p{30pt}|p{30pt}|p{30pt}|}\hline
$A_{\nu\nu'}$&  $d_{x^2-y^2}$
& $d_{xz}$ & $d_{yz}$ & $d_{xy}$ & $d_{z^2}$\\ 
\hline $d_{x^2-y^2}$ &  0.13  & 0.12  & 0.12  & 0.01 & 0.40 \\ 
\hline $d_{xz} $& 0.12  & 0.04 & 0.24 & 0.08 & 0.40  \\ 
\hline $d_{yz}$ & 0.12 & 0.24  & 0.04  & 0.08 & 0.40\\
\hline $d_{xy} $& 0.01 & 0.08  & 0.08 & 0 & 0.01 \\
\hline $d_{z^2}$ & 0.40  & 0.40  & 0.40   & 0.01 & 1.31  \\\hline
\end{tabular}
\end{table}

\begin{table}
\caption{Fourth-order sum of matrix elements 
contributing to the nnn spin-spin couplings.}
\label{table6}
\begin{tabular}{|p{30pt}|p{30pt}| p{30pt}|p{30pt}|p{30pt}|p{30pt}|}\hline
$A^{\prime}_{\nu\nu'}$&  $ d_{x^2-y^2}$
& $d_{xz}$ & $d_{yz}$ & $d_{xy}$ & $d_{z^2}$\\ 
\hline $d_{x^2-y^2}$ &  0.18  & 0.46  & 0.46  & 0 & 0.42 \\ 
\hline $d_{xz} $& 0.46  & 4.66 & 0 & 0 & 0.53  \\ 
\hline $d_{yz}$ & 0.46 & 0  & 4.66  & 0 & 0.53\\
\hline $d_{xy} $& 0 & 0  & 0 & 0 & 0 \\
\hline $d_{z^2}$ & 0.42  & 0.53  & 0.53   & 0 & 1.32  \\\hline
\end{tabular}
\end{table}

\subsection{Hopping between same-type Fe orbitals}

There are contributions
to the  effective hopping matrix elements due to second order
processes. Namely, processes in which an
electron from a doubly occupied As orbital momentarily hops to
the nn Fe $d$ orbital and then an electron from the
doubly occupied doped $d$-orbital hops to the singly occupied
As orbital left behind.  These processes give rise to the following expression
\begin{eqnarray}
{\delta t}_{\nu,\nu}= \sum_{\alpha=1}^6
{{V^{\alpha\nu}_1 V^{\alpha\nu}_2} \over {U^* + \epsilon}},
\end{eqnarray}
where $\epsilon$ is given by Eq.~\ref{epsilon}.
Again, since the value of $U^*$ is not known, for large enough values
of $U^*$ ($\epsilon < 1 eV$) we can neglect $\epsilon$ in the above
expression to obtain the following expression
\begin{eqnarray}
{\delta t}_{\nu,\nu}&=& {{B_{\nu\nu}} \over {U^*}}, \\
B_{\nu\nu} &=& \sum_{\alpha=1}^6 V^{\alpha\nu}_1 V^{\alpha\nu}_2,
\end{eqnarray}
and the values of $B_{\nu\nu}$ are given as the diagonal
elements in in Table~\ref{table7}. The actual
estimates for $\delta t_{\nu\nu}$ can be obtained by dividing 
the values in the table  by $U^*$.
The total effective nn hopping is given as
\begin{eqnarray}
{\tilde t}_{\nu\nu} = t_{\nu\nu} + \delta t_{\nu\nu}.
\end{eqnarray}
In the case of next nn such as the Fe atoms 1 and 3 in
Fig~\ref{fig1} we obtain
\begin{eqnarray}
{\tilde t}'_{\nu\nu}&=& {{B'_{\nu\nu}} \over {U^*}}, \\
B'_{\nu\nu} &=& {1 \over 2} \sum_{\alpha=1}^6 (V^{\alpha\nu}_1)^2.
\end{eqnarray}
The values of $B'_{\nu\nu}$ are given as the diagonal elements of
Table~\ref{table8}.

\begin{table}
\caption{Second-order terms contributing to the 
effective nn hopping.}
\label{table7}
\begin{tabular}{|p{30pt}|p{30pt}| p{30pt}|p{30pt}|p{30pt}|p{30pt}|}\hline
$B_{\nu\nu'} $&  $d_{x^2-y^2}$
& $d_{xz}$ & $d_{yz}$ & $d_{xy}$ & $d_{z^2}$\\ 
\hline $d_{x^2-y^2}$ &  -0.50  & 0.31  & 0.31  & 0 & 0 \\ 
\hline $d_{xz} $& 0.31  & 0.29 & -0.49 & -0.29 & -0.65  \\ 
\hline $d_{yz}$ & 0.31 & -0.49  & 0.29  & 0.29 & 0.65\\
\hline $d_{xy} $& 0 & -0.29  & 0.29 & 0 & 0.1 \\
\hline $d_{z^2}$ & 0  & -0.65  & 0.65   & 0.01 & 1.62  \\\hline
\end{tabular}
\end{table}

\begin{table}
\caption{Second-order terms contributing to the 
effective nnn hopping.}
\label{table8}
\begin{tabular}{|p{30pt}|p{30pt}| p{30pt}|p{30pt}|p{30pt}|p{30pt}|}\hline
$B^{\prime}_{\nu\nu'} $& $d_{x^2-y^2}$  
& $d_{xz}$ & $d_{yz}$ & $d_{xy}$  & $d_{z^2}$\\ 
\hline $d_{x^2-y^2}$ &  0.82  & -0.93  & -0.93  & 0 & 0 \\ 
\hline $d_{xz} $& -0.93  & 2.60 & 0 & -0.02 & -0.99  \\ 
\hline $d_{yz}$ & -0.93 & 0  & 2.60  & 0.02 & 0.99\\
\hline $d_{xy} $& 0 & -0.02  & 0.02 & 0.08 & 0 \\
\hline $d_{z^2}$ & 0  & -0.99  & 0.99   & 0 & 1.74  \\\hline
\end{tabular}
\end{table}

\subsection{Hopping between different-type Fe orbitals}

There is a second order process by means of which 
a doubly occupied site can effectively hop to a nn
Fe $d$ orbital of different type by involving an 
intervening As $sp$ orbital. These contributions
are smaller than those connecting two of the same-type Fe $d$
orbitals and they can be approximated by
\begin{eqnarray}
{\delta t}_{\nu,\nu'}&=& {{B_{\nu\nu'}} \over {U^*}}, \\
B_{\nu\nu'} &=& \sum_{\alpha=1}^6 V^{\alpha\nu}_1 V^{\alpha\nu'}_2.
\end{eqnarray}
The values of $B_{\nu\nu'}$ are given by the off-diagonal elements
of Table~\ref{table7}.

There is also a second order process which gives rise to hopping
between nnn Fe $d$ orbitals of different type. For this
case we obtain
\begin{eqnarray}
{\delta t}'_{\nu,\nu'}&=& {{B'_{\nu\nu'}} \over {U^*}} \\
B'_{\nu\nu'} &=& \sum_{\alpha=1}^6 V^{\alpha\nu}_1 V^{\alpha\nu'}_1.
\end{eqnarray}
The values of $B'_{\nu\nu'}$ are given by the off-diagonal elements
of Table~\ref{table8}.

The total effective nn and nnn hopping for $\nu\ne\nu'$ are given as
\begin{eqnarray}
{\tilde t}_{\nu\nu'} &=& t_{\nu\nu'} + \delta t_{\nu\nu'}, \\
{\tilde t}'_{\nu\nu'} &=& \delta t'_{\nu\nu'}.
\end{eqnarray}

\subsection{Effective Hamiltonian}

In summary the effective hamiltonian is given by
\begin{eqnarray}
{\cal H}_{eff} &=& \sum_{\nu,\nu'}{\cal H}_{\nu\nu'}
 - J^H \sum_{i,\mu,\mu'} \vec S^{\mu}_i\cdot \vec S^{\mu'}_i,
\label{effective0}
\end{eqnarray}
where each of the ${\cal H}_{\nu,\nu'}$ terms above may be written as
\begin{eqnarray}
{\cal H}_{\nu,\nu'} & = & - \sum_{<ij>,\sigma} {\tilde t}_{\nu,\nu'}
c^{\dagger}_{j\nu'\sigma}c_{i\nu\sigma}  -
\sum_{<<ij>>,\sigma} {\tilde t}^{\prime}_{\nu,\nu'}
c^{\dagger}_{j\nu'\sigma}c_{i\nu\sigma} \nonumber \\
&+& \sum_{<ij>} J_{\nu,\nu'}
\vec S^{\nu}_i \cdot \vec S^{\nu'}_j
+ \sum_{<<ij>>} J^{\prime}_{\nu,\nu'}
\vec S^{\nu}_i \cdot \vec S^{\nu'}_j.
\label{effective}
\end{eqnarray}

Next, we will provide estimates of the coupling constants
${\tilde t}_{\nu,\nu'}$, ${\tilde t}'_{\nu,\nu'}$, ${J}_{\nu,\nu'}$
and ${J}'_{\nu,\nu'}$ involved in the above model
based on the values of
the parameters obtained from fitting the LDA results to
the tight binding model.
The matrix elements for
nn hopping ${\tilde t}_{\nu\nu'}$ and spin-spin interaction
$J_{\nu\nu'}$ as well as their counterparts for next nn
interactions, that is, between sites diagonally across in the square
lattice formed by the Fe atoms,  $\tilde t^{\prime}_{\nu\nu'}$ and
$J^{\prime}_{\nu\nu'}$, are given
in  Tables~\ref{table9},\ref{table10},
for $U^*=U= 5$ eV and in Tables~\ref{table11},\ref{table12} 
using  $U^*_{\nu}=U=3$ eV.
A more simplified model than the one given above is discussed in the following 
section.

\begin{table}
\caption{The estimated matrix elements for the $J_{\nu\nu'}$
and $J'_{\nu\nu'}$ for $U=5$ eV.}\label{table9}
\begin{tabular}{|p{30pt}|p{30pt}| p{30pt}|p{30pt}|p{30pt}|p{30pt}|}\hline
$ J $ & $d_{x^2-y^2}$  & $d_{xz}$ & $d_{yz}$ & $d_{xy}$ & $d_{z^2}$\\ 
$ J'$ &  &  &  &  & \\ 
\hline $d_{x^2-y^2}$ &   0.04 & 0.0 & 0.0 & 0.0 & 0.01 \\
  &   0 & 0 & 0.01 & 0.01 & 0.01 \\
\hline $d_{xz} $&  0.0 & 0.20 & 0.24 & 0.0 & 0.01\\
&  0.01 & 0.07 & 0 & 0.0 & 0.01\\
\hline $d_{yz}$ &  0.0 & 0.24 & 0.20 & 0.0 & 0.01 \\
 &  0.01 & 0 & 0.07 & 0.0 & 0.01\\
\hline $d_{xy} $ & 0.0 &  0.0 & 0.0 & 0.15 & 0.03 \\
 & 0 & 0 & 0 & 0 & 0\\
\hline $d_{z^2}$ & 0.01 & 0.01 & 0.01 & 0.03 & 0.08 \\
 &  0.01 & 0.01 & 0.01 & 0.0 & 0.02 \\ \hline
\end{tabular}
\end{table}

\begin{table}
\caption{The estimated matrix elements for the $t_{\nu\nu'}$
and $t'_{\nu\nu'}$
 for $U=5$ eV.}\label{table10}
\begin{tabular}{|p{30pt}|p{30pt}| p{30pt}|p{30pt}|p{30pt}|p{30pt}|}\hline
$ {\tilde t} $ &   $d_{x^2-y^2}$
& $d_{xz}$ & $d_{yz}$ & $d_{xy}$ & $d_{z^2}$\\ 
$ {\tilde t}'$ & & & &  & \\ 
\hline $d_{x^2-y^2}$ &  0.12  & 0.06  & 0.06  & 0 & 0 \\ 
                 &  0.08  & -0.09  & -0.09  & 0 & 0 \\ 
\hline $d_{xz} $& 0.06  & 0.56 & 0.44 & -0.06 & -0.13  \\ 
                &  -0.09  &   0.26 &  0 & 0 & -0.1  \\ 
\hline $d_{yz}$ & 0.06 & 0.44  & 0.56  & 0.06 & 0.13\\
                 & -0.09 & 0  & 0.26  & 0 & 0.1\\
\hline $d_{xy} $& 0  & -0.06  & 0.06 & 0.43  & 0.22 \\
               &  0     &  0      &   0   &  0.01 & 0 \\
\hline $d_{z^2}$ & 0  & -0.13  & 0.13  & 0.22 & 0.54   \\
                 & 0  & -0.1  & -0.1  & 0 & 0.17   \\\hline
\end{tabular}
\end{table}
 
\begin{table}
\caption{The estimated matrix elements for the $J_{\nu\nu'}$
and $J'_{\nu\nu'}$ for $U=3$ eV.}\label{table11}
\begin{tabular}{|p{30pt}|p{30pt}| p{30pt}|p{30pt}|p{30pt}|p{30pt}|}\hline
$ J $ & $d_{x^2-y^2}$  & $d_{xz}$ & $d_{yz}$ & $d_{xy}$ & $d_{z^2}$\\ 
$ J'$ &  &  &  &  & \\ 
\hline $d_{x^2-y^2}$ &  0.08  & 0.02  & 0.02  & 0 & 0.06 \\ 
  &  0.01  & 0.03  & 0.03  & 0 & 0.03 \\ 
\hline $d_{xz} $& 0.02 & 0.34 & 0.42  & 0.01 & 0.06  \\ 
& 0.03 & 0.35 & 0  & 0 & 0.04  \\ 
\hline $d_{yz}$ & 0.02 & 0.42  & 0.34  & 0.01 & 0.06\\
 & 0.03 & 0  & 0.35  & 0 & 0.04\\
\hline $d_{xy} $& 0 &  0.01 & 0.01 & 0.25 & 0.05 \\
& 0 &  0 & 0 & 0 & 0 \\
\hline $d_{z^2}$ & 0.06  & 0.06  & 0.06   & 0.05  & 0.26  \\
 & 0.03  & 0.04  & 0.04   & 0 & 0.10  \\\hline
\end{tabular}
\end{table}

\begin{table}
\caption{The estimated matrix elements for the $t_{\nu\nu'}$
and $t'_{\nu\nu'}$
 for $U=3$ eV.}\label{table12}
\begin{tabular}{|p{30pt}|p{30pt}| p{30pt}|p{30pt}|p{30pt}|p{30pt}|}\hline
$ {\tilde t} $ &  $d_{x^2-y^2}$
& $d_{xz}$ & $d_{yz}$ & $d_{xy}$ & $d_{z^2}$\\ 
$ {\tilde t}'$ & & & &  & \\ 
\hline $d_{x^2-y^2}$ &  0.05 & 0.10 & 0.10 & 0 & 0\\
                 &   0.14 & -0.16 &-0.16 & 0 & 0\\
\hline $d_{xz} $& 0.10 & 0.60 & 0.38 & -0.10 &-0.22\\
                & -0.16 & 0.43 & 0 & 0 & -0.17\\
\hline $d_{yz}$ &  0.10 & 0.38 & 0.60 & 0.10 & 0.22 \\
                 &  -0.16 & 0 & 0.43 & 0 & 0.17\\
\hline $d_{xy} $&  0 & -0.10 & 0.10  & 0.43 & 0.23\\
               &   0 & 0 & 0  &  0.01 & 0\\
\hline $d_{z^2}$ &  0 &-0.22 & 0.22 &0.23 & 0.76\\
                 &  0 &-0.17 & 0.17 & 0 & 0.29\\
\hline
\end{tabular}
\end{table}

\section{Discussion}

First, by examining the 
Tables~\ref{table9},\ref{table10},\ref{table11},\ref{table12}, 
we notice that to a reasonable degree of approximation
the following three subspaces couple with each other rather weakly:
 (a)  one spanned by the $d_{x^2-y^2}$ Fe
orbital; (b) one spanned by 
the degenerate atomic Fe orbitals $d_{xz}$ and $d_{yz}$;  
(c) one spanned by the atomic Fe orbitals
$d_{xy}$ and $d_{z^2}$. Notice that the most significant off-diagonal
matrix elements are those which couple the $d_{xz}$ and
the $d_{yz}$ orbital and those  which couple the 
$d_{xy}$ to the $d_{z^2}$ orbital.
There are other smaller off-diagonal matrix elements
which couple these subspaces weakly. 
These three subspaces, however, are much more strongly 
coupled through $J^H$.

\begin{figure}[htp]
\vskip 0.4 in
\includegraphics[width=3.5 in]{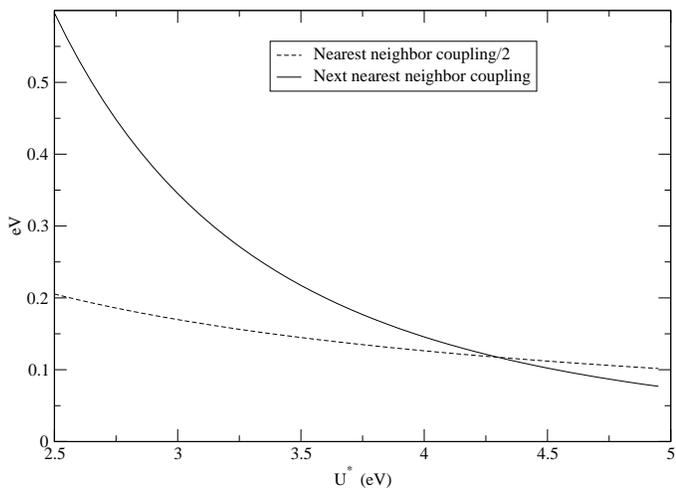}
\caption{The calculated  nn $J_{xz,xz}/2$
 is compared to the nnn $J'_{xz,xz}$ as a function of $U^*$ (using $U^*=U$).
The condition for the instability of the $(\pi,\pi)$ order
to the columnar order is $J'_{xz,xz}>J_{xz,xz}/2$
which occurs for values of $U^* \leq 4.3$ eV.}
\label{fig6}
\vskip 0.2 in
\end{figure}

We consider the undoped (LaOFeAs) case first. Because six electrons should
occupy the five Fe $d$ orbitals (see Fig.~\ref{fig5}), the lowest energy 
subspace
spanned by $d_{x^2-y^2}$ is occupied by two electrons, and also 
each of the  other 
two subspaces is also occupied by two electrons.
As mentioned above, these three subspaces are coupled mainly
because $J^H\ne0$. Furthermore, the bands formed in any given subspace 
are intersected by the bands formed in the
other two subspaces because their atomic energy difference
is small compared to their bandwidth.

There is a great degree of magnetic
frustration, as noted in Refs.~[\onlinecite{Sachdev,AF,Yildirim}],
especially in the subspace spanned by $d_{xz}/d_{yz}$. In this
subspace the next nn coupling $J'$ becomes greater than $J/2$
 for any value of $U^*\leq  4.3 eV$ (using $U^*=U$)(See Fig.~\ref{fig6}).
When $J<2 J'$, the  observed
columnar antiferromagnetic ordering
is favorable\cite{Sachdev,AF} relative to the familiar
$(\pi,\pi)$ antiferromagnetic order.
The subspace spanned by $d_{x^2-y^2}$ has net spin zero and, therefore,
is not expected to contribute significantly to the choice of magnetic order.
On the other hand, there seems to be less frustration in the 
subspace spanned by $d_{xy}/d_{z^2}$, which is half-filled;
we therefore expect that a long-range antiferromagnetic order
should characterize the ground state of this subspace if it
were uncoupled from the $d_{xz}/d_{yz}$ subspace. However, due to Hund's
rule coupling $J^H$, the spin orientation of all subspaces
should be common. The conflicting preferences
of these two subspaces, which are forced to make a common choice,
introduces  further frustration of relative spin orientation.

We expect that the subspace spanned by $d_{xz}/d_{yz}$ drives
the system to a global columnar  order\cite{AF,Sachdev,Yildirim} 
because it is characterized by the larger couplings.
The presence of a large $J'$ in the subspace spanned by $d_{xz}/d_{yz}$  
might impose the observed columnar  order through the relatively large
Hund's rule coupling $J_H \sim 0.5 eV$. 
The fact that the subspace $d_{xz}/d_{yz}$
prefers the columnar order and the subspace $d_{z^2}/d_{xy}$ 
prefers the $(\pi,\pi)$ order creates frustration which may
also explain the fact that the observed moment
per Fe atom is small.

We emphasize that unlike the case of undoped cuprates, the
undoped parent compound in the case of the oxypnictides
is not an insulator. As can be inferred from Fig.~\ref{fig1}, the motion
along the ferromagnetic direction is not hindered and, therefore,
the undoped material is expected to demonstrate anisotropic
transport in the SDW phase.

We would like to discuss the case of the neutral FeAs layer
which was considered in Sec.~\ref{LDA}, where it was found that the
ground state is characterized by $(\pi,\pi)$ order.
The case of the neutral layer has five electrons per Fe atom and
this implies that all Fe $d$ orbitals should be singly
occupied.  Therefore, the subspace 
spanned by $d_{x^2-y^2}$  is no longer
characterized by spin zero. This means that the subspace
$d_{xz}/d_{yz}$ in order to drive the columnar order has 
to compete against not just one but two subspaces which prefer 
the $(\pi,\pi)$ order. 

The phenomenological hamiltonian considered in
Ref.~[\onlinecite{AF,Yildirim}] and in Ref.~[\onlinecite{Sachdev}] to
introduce frustration, is different from  the one we derived based
on a more rigorous approach, which is more complex.
The next step would be to study the hamiltonian given in Eqs. 
(\ref{effective0}),(\ref{effective})
by various analytical and numerical techniques, which is beyond the 
scope of the  present work.

While the estimated nn antiferromagnetic coupling constants are
of similar magnitude to the one in the cuprous oxides\cite{RMP},
in the oxypnictide materials there is magnetic frustration mainly 
due to the fact that
the nnn antiferromagnetic coupling for the $d_{xz}/d_{yz}$ subspace is
large. Therefore, assuming that the pairing interaction 
between electrons is of magnetic origin, it is not clear if the
pairing energy scale is larger or smaller compared
to that in the cuprate superconducting materials. The 
pairing energy scale in the present model may be enhanced by the
``flavor'' factor, that is, the number of states spanning the subspace where
the added electrons go in the case of electron doping,
and by the fact that the hopping and spin-exchange matrix elements
are estimated to be somewhat larger compared to those in the 
case of the cuprates (see
Tables~\ref{table9},\ref{table10},\ref{table11},\ref{table12}).
Therefore, it is conceivable that this new class
of superconductors could
lead to higher critical temperatures upon future optimization
of the doping agents and other factors.

A very important difference between the oxypnictides and the
cuprates is that the five-fold sector can be thought of
as formed by three subsectors,
two 2-flavor sectors and a third 1-flavor sector.
The $d_{xz}/d_{yz}$ sector prefers the SDW order depicted in Fig.~\ref{fig1}, 
the sector spanned by the $d_{x^2-y^2}$ orbital has spin zero and 
the other sector spanned by $d_{xy}/d_{z^2}$
prefers antiferromagnetic long-range
order. These subspaces are coupled by Hund's rule which, we believe,
leads to the SDW order with ferromagnetic order along one
direction and antiferromagnetic ordering between such chains.
As in the case of cuprates, 
superconductivity in the oxypnictide materials might coexist with SDW order\cite{Drew}
but these are expected to be to some extent competing orders as found
in neutron\cite{Qiu} and $\mu SR$ studies\cite{Carlo} done on the
superconducting doped materials.

 \acknowledgements{
We would to thank  B. Halperin, E. Demler and S. Sachdev for
useful discussions and  C. Xu  for useful comments on
the manuscript.
}

\end{document}